\documentclass[pra,twocolumn,superscriptaddress,showpacs]{revtex4}

\def\bra#1{\mathinner{\langle{#1}|}}
\def\ket#1{\mathinner{|{#1}\rangle}}
\def\braket#1{\mathinner{\langle{#1}\rangle}}

\let\protect\relax
{\catcode`\|=\active
  \xdef\Braket{\protect\expandafter\noexpand\csname Braket \endcsname}
  \expandafter\gdef\csname Braket \endcsname#1{\begingroup
     \ifx\SavedDoubleVert\relax
       \let\SavedDoubleVert\|\let\|\BraDoubleVert
     \fi
     \mathcode`\|32768\let|\BraVert
     \left\langle{#1}\right\rangle\endgroup}
}
\def\BraVert{\@ifnextchar|{\|\@gobble}
     {\egroup\,\mid@vertical\,\bgroup}}
\def\BraDoubleVert{\egroup\,\mid@dblvertical\,\bgroup}
\let\SavedDoubleVert\relax

{\catcode`\|=\active
  \xdef\set{\protect\expandafter\noexpand\csname set \endcsname}
  \expandafter\gdef\csname set \endcsname#1{\mathinner
        {\lbrace\,{\mathcode`\|32768\let|\midvert #1}\,\rbrace}}
  \xdef\Set{\protect\expandafter\noexpand\csname Set \endcsname}
  \expandafter\gdef\csname Set \endcsname#1{\left\{%
     \ifx\SavedDoubleVert\relax \let\SavedDoubleVert\|\fi
     \:{\let\|\SetDoubleVert
     \mathcode`\|32768\let|\SetVert
     #1}\:\right\}}
}
\def\midvert{\egroup\mid\bgroup}
\def\SetVert{\@ifnextchar|{\|\@gobble}
    {\egroup\;\mid@vertical\;\bgroup}}
\def\SetDoubleVert{\egroup\;\mid@dblvertical\;\bgroup}

\begingroup
 \edef\@tempa{\meaning\middle}
 \edef\@tempb{\string\middle}
\expandafter \endgroup \ifx\@tempa\@tempb
 \def\mid@vertical{\middle|}
 \def\mid@dblvertical{\middle\SavedDoubleVert}
\else
 \def\mid@vertical{\mskip1mu\vrule\mskip1mu}
 \def\mid@dblvertical{\mskip1mu\vrule\mskip2.5mu\vrule\mskip1mu}
\fi

\usepackage{graphicx}
\usepackage[T1]{fontenc}
\usepackage[]{amsfonts}
\usepackage[]{amsmath}
\usepackage[]{rotating}
\usepackage[]{color}
\usepackage[]{float}
\usepackage[]{fancyhdr}
\usepackage[]{booktabs}
\usepackage[]{epsfig}
\usepackage{appendix}
\usepackage{amssymb}
\usepackage{psfrag}
\usepackage{setspace}

\begin{document}

\title{Adiabatically steered open quantum systems: Master equation and optimal phase}

\author{J. Salmilehto}
\affiliation{Department of Applied Physics/COMP, Aalto University, P.O. Box 14100, FI-00076 AALTO, Finland}
\author{P. Solinas}
\affiliation{Department of Applied Physics/COMP, Aalto University, P.O. Box 14100, FI-00076 AALTO, Finland}
\author{J. Ankerhold}
\affiliation{Institut für Theoretische Physik, Universität Ulm, Albert-Einstein-Allee 11, 89069 Ulm, Germany}
\author{M. Möttönen}
\affiliation{Department of Applied Physics/COMP, Aalto University, P.O. Box 14100, FI-00076 AALTO, Finland}
\affiliation{Low Temperature Laboratory, Aalto University, P.O. Box 13500, FI-00076 AALTO, Finland}

\pacs{03.65.Vf, 03.65.Yz, 05.30.-d}

\begin{abstract}

We introduce an alternative way to derive the generalized form of
the master equation recently presented in
Ref.~[J.~P.~Pekola~\textit{et
al}.,~Phys.~Rev.~Lett.~\textbf{105},~030401~(2010)] for an
adiabatically steered two-level quantum system interacting with a
Markovian environment. The original derivation employed the
effective Hamiltonian in the adiabatic basis with the standard
interaction picture approach but without the usual secular
approximation. Our approach is based on utilizing a master equation
for a non-steered system in the first super-adiabatic basis. It is
potentially efficient in obtaining higher-order equations.
Furthermore, we show how to select the phases of the adiabatic
eigenstates to minimize the local adiabatic parameter and how this
selection leads to states which are invariant under a local gauge
change. We discuss also the effects of the adiabatic noncyclic
geometric phase on the master equation.

\end{abstract}

\maketitle

\section{\label{sec:intro}Introduction}

The adiabatic theorem \cite{zp51/165, psj5/435} has been one of the workhorses of quantum physics for decades. It states that if the external control parameters of the system Hamiltonian vary slowly enough, the system remains very accurately in one of its initial instantaneous eigenspaces. As slowly varying quantum systems appear in many fields of physics, a multitude of applications for the theorem exists. In recent years, adiabatically steered quantum systems have attracted a lot of interest due to their connection to geometric phases in cyclic evolution \cite{prsla392/45, prl51/2167, prl52/24}. These phases provide a potential alternative to quantum information processing \cite{QCaQI, PoQCaI, QC_Nakahara} in which the quantum gates are implemented by purely geometric means \cite{pla264/94, pra61/010305(R), pra64/022310, pra59/2910, nature403/869, nature407/355, science292/1695, pra66/022102, prl90/028301, pra67/062315}. This \textit{geometric quantum computation} has been shown to offer inherent robustness against control errors \cite{pra65/012322, prl91/090404, pra70/042316, pra72/020301(R)} due to the fact that geometric phases depend only on some global geometric properties. Different ways to describe geometric phases in open systems have been introduced \cite{prl85/2845, pra67/020101(R), pra71/012331, prl95/250503, pra73/062101, pra72/022328} and methods to account for the effect of the environment on the system evolution have been studied \cite{pra65/012322, prl90/160402, prl92/020402, prl94/020503, prl94/070407, pra73/052304, pra73/022327, pra76/012337, prb77/115322} along with techniques to reduce the unwanted noise. However, a consistent description of the combined effect of adiabatic steering and noise was missing until recently, when a master equation was introduced in Refs.~\cite{master, general_master}.

In the approach of Ref.~\cite{master}, it was shown that the
typically applied secular approximation \cite{API, prl94/070407} is
not suitable in describing adiabatic evolution. Taking into account
all the relevant contributions leads to a master equation which
ensures relaxation to a proper basis and shows that the ground-state
dynamics are not influenced by zero-temperature Markovian noise in
the adiabatic limit. Thus the system exhibits inherent robustness.
The master equation derived in Ref.~\cite{master} was generalized to
hold for a generic system--environment coupling operator in
Ref.~\cite{general_master}. Furthermore, the master equation was
applied to describe Cooper pair pumping \cite{prb73/214523,
prl100/177201, apl90/082102} in Refs.~\cite{master, general_master}.

In this paper, we introduce an alternative derivation of the master
equation for adiabatically steered quantum systems coupled to a
Markovian environment. Our derivation is based on utilizing a
non-steered master equation in the first super-adiabatic basis. We
show that the master equation we obtain is the same as in
Ref.~\cite{general_master}. Our method is potentially more efficient
in obtaining higher order expansions in the adiabatic parameter. In
addition, we introduce a way to select the complex phases of the
adiabatic basis states such that the local adiabatic parameter is
minimized leading to vanishing diagonal elements for the operator
describing the steering. We show that this selection results in
locally phase invariant basis states. Finally, we discuss how to
account for the time-local accumulation speed of the geometric phase
in the environment-induced transitions.

The structure of this paper is as follows. In the next section, we introduce our model describing the open quantum system. In Sec.~\ref{sec:non-steered}, we derive the master equation for a non-steered system subject to decoherence. In Sec.~\ref{sec:full_master}, we use the non-steered master equation to obtain the full master equation for adiabatic steering. In Sec.~\ref{sec:opt_phase}, we introduce the optimal phase selection for the adiabatic eigenstates and demonstrate the main implications of such a selection. We conclude the paper in Sec.~\ref{sec:conclusions}.

\section{\label{sec:model}Model}

We consider a quantum system with a Hamiltonian $\hat{H}_S$ which depends on a set of real control parameters $\{q_k\}$ that vary in time. The system is assumed to be interacting with the environment so that the total Hamiltonian is
\begin{equation}
\hat{H}(t) = \hat{H}_S(t) + \hat{V}(t) + \hat{H}_E,
\label{eq:H}
\end{equation}
where $\hat{V}(t)$ is the coupling between the system and its environment and $\hat{H}_E$ is the Hamiltonian of the environment. We assume that the coupling is of the generic form $\hat{V} = \hat{A} \otimes \hat{X}(t)$, where $\hat{A}$ is the system part of the coupling operator and $\hat{X}(t)$ acts in the Hilbert space of the environment. Let $\ket{m;\vec{q}(t)}$ be the instantaneous eigenstate of $\hat{H}_S(t)$ and $E_m(t)$ the corresponding eigenenergy defined by $\hat{H}_S[\vec{q}(t)] \ket{m;\vec{q}(t)} = E_m[\vec{q}(t)]\ket{m;\vec{q}(t)}$. In the context of adiabatic evolution, $\{\ket{m;\vec{q}(t)}\}$ is referred to as the \textit{adiabatic basis}. We assume that the adiabatic states are normalized and non-degenerate.

Let the Hamiltonian $\hat{H}_S(t)$ be diagonalized in a fixed basis $\{\ket{m_f}\}$ using the eigendecomposition as $\hat{\tilde{H}}_S(t)=\hat{D}^{\dagger}(t)\hat{H}_S(t)\hat{D}(t)$, implying that $\braket{n_f|\hat{\tilde{H}}_S(t)|m_f} = E_m(t) \delta_{nm}$. We define a similar transformation for the total density operator $\hat{\rho}(t)$ in the Schrödinger picture as $\hat{\tilde{\rho}}(t)=\hat{D}^{\dagger}(t)\hat{\rho}(t)\hat{D}(t)$. It follows from the Schrödinger equation that the evolution of $\hat{\tilde{\rho}}(t)$ is governed by the effective Hamiltonian for the adiabatic basis
\begin{equation}
\hat{\tilde{H}}^{(1)}(t) = \hat{\tilde{H}}_S(t) + \hbar \hat{w}(t) + \hat{\tilde{V}}(t) + \hat{H}_E,
\label{eq:Heff}
\end{equation}
where $\hat{\tilde{V}}(t)=\hat{D}^{\dagger}(t)\hat{V}(t)\hat{D}(t)=\hat{D}^{\dagger}(t)\hat{A}\hat{D}(t) \otimes \hat{X}(t)$ and $\hat{w}(t) = -i\hat{D}^{\dagger}(t)\dot{\hat{D}}(t)$. Omitting the environment and assuming adiabatic evolution, a more accurate approximation for the exact evolving state is achieved if the adiabatic states are corrected by
\begin{equation}
\ket{\delta m;\vec{q}(t)} = -i\hbar \sum_{k \neq m}\ket{k;\vec{q}(t)}\frac{\braket{k;\vec{q}(t)|\frac{\partial}{\partial t}|m;\vec{q}(t)}}{E_{m}-E_{k}},
\label{eq:delta_m}
\end{equation}
in the first order in the perturbation theory. The basis formed by the corrected states $\{\ket{m}+\ket{\delta m}\}$ is usually referred to as the \textit{first superadiabatic basis} \cite{prsla392/45}.

We introduce the local adiabatic parameter as $\alpha(t) = \hbar
||\hat{w}(t)||/\Delta(t)$, where we compare the Hilbert-Schmidt norm
of the operator arising from the adiabatic evolution $||\hat{w}(t)||
= \sqrt{\textrm{Tr}_S\{\hat{w}(t)^{\dagger}\hat{w}(t)\}}$ to an
instantaneous minimum energy gap in the spectrum $\Delta(t)$. Here
$\textrm{Tr}_S$ denotes the trace over the system degrees of freedom
and in the following we will use $\mathrm{Tr}_E$ to denote the trace
over the environment degrees of freedom. The parameter $\alpha(t)$
yields typically a good estimate for the degree of adiabaticity of
the evolution \cite{pra71/012331, general_master}. In cyclic
evolution with the period $T$, the parameter scales as $1/T$ and,
thus, in adiabatic evolution we should have $\alpha(t) \ll 1$.

\section{\label{sec:non-steered}Master equation for a non-steered system}

Let us study the dynamics of a generic non-steered two-level quantum system coupled to its environment. Denote the ground and excited states of $\hat{H}_S$ in the Schrödinger picture as $\ket{g}$ and $\ket{e}$, respectively, with corresponding eigenenergies $E_g$ and $E_e$. We apply the standard method \cite{master, general_master, API} to obtain the master equation for a non-steered system as
\begin{equation}
\frac{d\rho_{gg}}{dt} = - (\Gamma_{ge}+\Gamma_{eg})\rho_{gg} + \Re e\{ \tilde{\Gamma}_0\rho_{ge} \} + \Gamma_{eg},
\label{eq:rho_gg1}
\end{equation}
and
\begin{equation}
\begin{split}
\frac{d\rho_{ge}}{dt} &= i\omega_{01}\rho_{ge} - (\tilde{\Gamma}_++\tilde{\Gamma}_-)\rho_{gg} \\ &- \bigg( \frac{\Gamma_{eg}}{2}+\frac{\Gamma_{ge}}{2}+\Gamma_{\varphi} \bigg) \rho_{ge} + (\Gamma_{\alpha}+\Gamma_{\beta})\rho_{eg} + \tilde{\Gamma}_+,
\end{split}
\label{eq:rho_ge}
\end{equation}
where $\rho_{rs} = \braket{r|\hat{\rho}_S|s}$ with $r,s \in \{g,e\}$, and $\omega_{01} = (E_e-E_g)/\hbar$. The transition rates are defined as
\begin{gather*}
\Gamma_{ge} = \frac{|\braket{e|\hat{A}|g}|^2}{\hbar^2} S_{X}(-\omega_{01}), \\
\Gamma_{eg} = \frac{|\braket{e|\hat{A}|g}|^2}{\hbar^2} S_{X}(+\omega_{01}), \\
\tilde{\Gamma}_0 = \frac{\braket{e|\hat{A}|g}( \braket{g|\hat{A}|g}-\braket{e|\hat{A}|e})}{\hbar^2} S_X(0), \\
\tilde{\Gamma}_{\pm}=\frac{\braket{g|\hat{A}|e}(\braket{e|\hat{A}|e}-\braket{g|\hat{A}|g})}{2\hbar^2}S_X(\pm\omega_{01}), \\
\Gamma_{\varphi}=\bigg( \frac{|\braket{e|\hat{A}|e}|^2}{2\hbar^2} + \frac{|\braket{g|\hat{A}|g}|^2}{2\hbar^2} - \frac{\braket{g|\hat{A}|g}\braket{e|\hat{A}|e}}{\hbar^2} \bigg) S_X(0), \\
\Gamma_{\alpha}=\frac{\braket{g|\hat{A}|e}^2}{2\hbar^2}S_X(\omega_{01}), \\
\Gamma_{\beta}=\frac{\braket{g|\hat{A}|e}^2}{2\hbar^2}S_X(-\omega_{01}).
\end{gather*}
The spectral density is denoted by $S_X(\omega) = \int_{-\infty}^{\infty} d\tau \mathrm{Tr}_E \{\hat{\rho}_E \hat{X}(\tau) \hat{X}(0)\} e^{i\omega \tau}$. Note that we neglect the drive, i.e., omit all terms proportional to $\hat{w}$. Furthermore, Eqs.~(\ref{eq:rho_gg1}) and (\ref{eq:rho_ge}) include all the nonsecular terms neglected in the usual application of the approach of Ref.~\cite{API}.

For details concerning the derivation of Eqs.~(\ref{eq:rho_gg1}) and (\ref{eq:rho_ge}) see Appendix. Especially, we neglect the possible imaginary parts of the transition rates, i.e., the Lamb shift, by assuming that the system time scales are longer than the system autocorrelation time.

\section{\label{sec:full_master}Master equation for adiabatic steering}

We aim to derive the the full master equation for the system coupled to its environment in adiabatic steering using the master equation for a non-steered system. Define a unitary transformation $\hat{D}_1(t)$ making $\hat{\tilde{H}}_S(t) + \hbar \hat{w}(t)$ diagonal in the fixed basis $\{ \ket{0}, \ket{1}\}$. Thus the evolution of the density matrix $\hat{\tilde{\rho}}^{(2)} = \hat{D}_1^{\dagger}\hat{\tilde{\rho}}\hat{D}_1 = \hat{D}_1^{\dagger}\hat{D}^{\dagger}\hat{\rho}\hat{D}\hat{D}_1$ is governed by the effective Hamiltonian for the first super-adiabatic basis
\begin{equation}
\hat{\tilde{H}}^{(2)}(t) = \hat{\tilde{H}}^{(2)}_S(t) + \hbar \hat{w}_1(t) + \hat{\tilde{V}}^{(2)}(t) + \hat{H}_E,
\label{eq:Heff2}
\end{equation}
where $\hat{\tilde{H}}^{(2)}_S(t)=\hat{D}_1^{\dagger}(t)[\hat{\tilde{H}}_S(t)+\hbar \hat{w}(t)]\hat{D}_1(t)$, $\hat{\tilde{V}}^{(2)}(t)=\hat{D}_1^{\dagger}(t)\hat{\tilde{V}}(t)\hat{D}_1(t)$, and $\hat{w}_1=-i\hat{D}_1^{\dagger}(t)\dot{\hat{D}}_1(t)$.

Assume that the super-adiabatic correction, $\hat{w}_1$, is negligible with respect to the adiabatic one so that we can write Eq.~(\ref{eq:Heff2}) as $\hat{\tilde{H}}^{(2)}(t) \approx \hat{\tilde{H}}^{(2)}_S(t) + \hat{\tilde{V}}^{(2)}(t) + \hat{H}_E$. Since this Hamiltonian describes effectively a non-steered system, we can employ the approach of Sec.~\ref{sec:non-steered} to write a master equation similar to Eqs.~(\ref{eq:rho_gg1}) and (\ref{eq:rho_ge}) as
\begin{equation}
\frac{d\rho_{gg}^{(2)}}{dt} = - (\Gamma_{ge}^{(2)}+\Gamma_{eg}^{(2)})\rho_{gg}^{(2)} + \Re e\{ \tilde{\Gamma}_0^{(2)}\rho_{ge}^{(2)} \} + \Gamma_{eg}^{(2)},
\label{eq:rho_gg1s}
\end{equation}
and
\begin{equation}
\begin{split}
\frac{d\rho_{ge}^{(2)}}{dt} &=  i\omega_{01}^{(2)}\rho_{ge}^{(2)} - (\tilde{\Gamma}_+^{(2)}+\tilde{\Gamma}_-^{(2)})\rho_{gg}^{(2)} \\ &- \bigg( \frac{\Gamma_{eg}^{(2)}}{2}+\frac{\Gamma_{ge}^{(2)}}{2}+\Gamma_{\varphi}^{(2)} \bigg) \rho_{ge}^{(2)} + (\Gamma_{\alpha}^{(2)}+\Gamma_{\beta}^{(2)})\rho_{eg}^{(2)} \\ &+ \tilde{\Gamma}_+^{(2)},
\end{split}
\label{eq:rho_ges}
\end{equation}
where we have marked the relevant terms in the super-adiabatic basis with the superscript $^{(2)}$ to avoid confusing them with the adiabatic ones. The transformation $\hat{D}_1(t)$ can be approximated using the perturbation theory for the adiabatic correction. This results in the superadiabatic eigenstates which we can obtain from Eq.~(\ref{eq:delta_m}) up to the linear order in $\alpha(t)$ in the two-state model as
\begin{equation}
\ket{g^{(2)}} = \ket{g}-\ket{e}\frac{w_{ge}^*}{\omega_{01}},
\label{eq:eig_corrected_g}
\end{equation}
and
\begin{equation}
\ket{e^{(2)}} = \ket{e}+\ket{g}\frac{w_{ge}}{\omega_{01}},
\label{eq:eig_corrected_e}
\end{equation}
with the eigenenergies $E_g^{(2)}=E_g+\hbar w_{gg}$ and $E_e^{(2)}=E_e+\hbar w_{ee}$, respectively. Here, we denote the matrix elements of the adiabatic correction as $w_{sr}=-i\braket{s|\dot{r}}$, where $r,s \in \{g,e\}$. Thus, the super-adiabatic energy gap up to this order becomes $\omega_{01}^{(2)}=\omega_{01}+(w_{ee}-w_{gg})$.

The matrix elements in Eqs.~(\ref{eq:rho_gg1s}) and
(\ref{eq:rho_ges}) can be written using the super-adiabatic
eigenstates to obtain the master equation for adiabatic steering up
to the linear order in $\alpha(t)$. We restrict our derivation to
the linear order since in the adiabatic limit, $\alpha(t)
\rightarrow 0$ making the contributions beyond the linear one
negligible. If we assume that the system is driven adiabatically but
does not necessarily remain in the ground state at all times, we
cannot assume the density matrix elements $\rho_{ge}$ to become
small enough to be neglected due to their order in this limit.
Hence, we are only considering $\alpha(t)$ as a small parameter and
neglect all terms with $\alpha^2$ or higher order. If we rewrite
$\rho_{gg}^{(2)}$ and $\rho_{ge}^{(2)}$ using
Eqs.~(\ref{eq:eig_corrected_g}) and (\ref{eq:eig_corrected_e}), the
master equation becomes
\begin{equation}
\begin{split}
& \dot{\rho}_{gg} - 2\frac{\Re e(w_{ge}^*\dot{\rho}_{ge})}{\omega_{01}} \\
&= - (\Gamma_{ge}^{(2)}+\Gamma_{eg}^{(2)})\left(\rho_{gg} - 2\frac{\Re e(w_{ge}^*\rho_{ge})}{\omega_{01}}\right) \\ &+ \Re e\left\{ \tilde{\Gamma}_0^{(2)} \left(\rho_{ge} + 2\frac{w_{ge}}{\omega_{01}}\rho_{gg} - \frac{w_{ge}}{\omega_{01}}\right)\right\} + \Gamma_{eg}^{(2)},
\end{split}
\label{eq:master_ggs_final}
\end{equation}
and
\begin{equation}
\begin{split}
& \dot{\rho}_{ge} + 2\frac{w_{ge}}{\omega_{01}}\dot{\rho}_{gg} \\
&= i[\omega_{01}+(w_{ee}-w_{gg})]\left( \rho_{ge} + 2\frac{w_{ge}}{\omega_{01}}\rho_{gg} - \frac{w_{ge}}{\omega_{01}} \right) \\ &- (\tilde{\Gamma}_+^{(2)}+\tilde{\Gamma}_-^{(2)})\left( \dot{\rho}_{gg} - 2\frac{\Re e(w_{ge}^*\dot{\rho}_{ge})}{\omega_{01}} \right) \\ &- \bigg( \frac{\Gamma_{eg}^{(2)}}{2}+\frac{\Gamma_{ge}^{(2)}}{2}+\Gamma_{\varphi}^{(2)} \bigg) \left( \rho_{ge} + 2\frac{w_{ge}}{\omega_{01}}\rho_{gg} - \frac{w_{ge}}{\omega_{01}} \right) \\ &+ (\Gamma_{\alpha}^{(2)}+\Gamma_{\beta}^{(2)}) \left( \rho_{eg} + 2\frac{w_{ge}^*}{\omega_{01}}\rho_{gg} - \frac{w_{ge}^*}{\omega_{01}}\right) + \tilde{\Gamma}_+^{(2)},
\end{split}
\label{eq:master_ges_final}
\end{equation}
where we have neglected all terms of order $\alpha^2$ or higher,
except in the $\Gamma^{(2)}$ terms which treat further below. We can
solve $\dot{\rho}_{gg}$ and $\dot{\rho}_{ge}$ from these equations
to obtain the full master equation. In addition, we employ
Eqs.~(\ref{eq:eig_corrected_g}) and (\ref{eq:eig_corrected_e}) to
rewrite the rates in the super-adiabatic approximation. To present
the full master equation, we adopt a notation which will not reduce
the generality of the equations but simplify them. In the nested
commutator expression for the master equation for a non-steered
system [see Eq.~(\ref{eq:APPdsigmadt_final}) in Appendix], the
coupling operator is only found in places where it is commuting with
other operators and, hence, provided that the Lamb shift is
neglected, we can add any operator comparable to the identity
operator to it without affecting the nested expression. Thus, the
system part of the coupling operator can be choosen traceless in the
two-state basis. We will adopt this convention by introducing
$m_1=\braket{g|\hat{A}|g}=-\braket{e|\hat{A}|e}$ and
$m_2=\braket{g|\hat{A}|e}$. Notice that $m_1 \in \mathbb{R}$ whereas
$m_2 \in \mathbb{C}$ in the case of a general coupling operator. The
master equation up to the linear order in $\alpha(t)$ and the
quadratic order in the system-environment coupling becomes
\begin{widetext}
\begin{equation}
\begin{split}
\dot{\rho}_{gg}   &= -2\Im m(w_{ge}^*\rho_{ge}) + S(\omega_{01})|m_2|^2 - [S(-\omega_{01})+S(\omega_{01})]|m_2|^2\rho_{gg} + 2[\Im m(m_2)\Im m(\rho_{ge})+\Re e(m_2)\Re e(\rho_{ge})]S(0)m_1 \\ &- 2\frac{2S(0)-S(-\omega_{01})-S(\omega_{01})}{\omega_{01}} \{ [\Im m(m_2)\Im m(w_{ge})+\Re e(m_2)\Re e(w_{ge})] [\Im m(m_2)\Im m(\rho_{ge})+\Re e(m_2)\Re e(\rho_{ge})] \} \\ &+ 2\frac{2S(0)-S(-\omega_{01})-S(\omega_{01})}{\omega_{01}} \{\Im m(m_2)\Im m(w_{ge})+\Re e(m_2)\Re e(w_{ge}) \} m_1\rho_{gg} \\ &-2\frac{S(0)-S(\omega_{01})}{\omega_{01}}m_1\{ \Im m(m_2)\Im m(w_{ge})+\Re e(m_2)\Re e(w_{ge}) \}
\end{split}
\label{eq:master_gg_complete}
\end{equation}
\end{widetext}
and
\begin{widetext}
\begin{equation}
\begin{split}
\dot{\rho}_{ge} &=  \ iw_{ge}(2\rho_{gg}-1)+i(w_{ee}-w_{gg})\rho_{ge}+i\omega_{01}\rho_{ge}-S(\omega_{01})m_1m_2 +[S(-\omega_{01})+S(\omega_{01})]m_1m_2\rho_{gg} -2S(0)m_1^2\rho_{ge} \\ &- i[S(-\omega_{01})+S(\omega_{01})]m_2[\Im m(\rho_{ge})\Re e(m_2)-\Im (m_2)\Re e(\rho_{ge})] \\ &- 2\frac{2S(0)-S(-\omega_{01})-S(\omega_{01})}{\omega_{01}} m_1^2w_{ge}\rho_{gg}  + 2\frac{S(0)-S(\omega_{01})}{\omega_{01}}m_1^2w_{ge} \\ &- im_2\frac{S(-\omega_{01})-S(\omega_{01})}{\omega_{01}}\{\Im m(m_2)\Re e(w_{ge})-\Im m(w_{ge})\Re e(m_2)\} \\ &- 2\frac{2S(0)-S(-\omega_{01})-S(\omega_{01})}{\omega_{01}}m_1 \{im_2[\Im m(w_{ge})\Re e(\rho_{ge})-\Im m(\rho_{ge})\Re e(w_{ge})] \\ &- [\Im m(m_2)\Im m(w_{ge})+\Re e(m_2)\Re e(w_{ge})]\rho_{ge}\}.
\end{split}
\label{eq:master_ge_complete}
\end{equation}
\end{widetext}

Here, we applied a shortened notation for the spectral densities
$S(\omega)=S_X(\omega)/\hbar^2$. We would like to emphasize that in
this section, we assume that the system is externally steered, i.e.,
the system Hamiltonian is time-dependent. Even though we do not
explicitly make the approximation of adiabatic rates \cite{API},
i.e., assume the evolution time is much longer than the environment
autocorrelation time so that $\omega_{01}$, $m_1$, $m_2$, and the
matrix elements of $\hat{w}$ vary slowly in time, the approximation
is implicitly assumed. This assumption stems from the fact that we
use the master equation for a non-steered system in the linear order
in $\alpha(t)$. Furthermore, we have neglected corrections
proportional to $\partial_{\omega} S(\omega) |_{\omega = \pm
\omega_{01}}$ when estimating the power spectra above since they
correspond to driving induced Lamb shift contributions which are
neglected also in the derivation in Refs.~\cite{master,
general_master}. In Sec.~\ref{sec:opt_phase}, we show that $w_{gg}$
and $w_{ee}$ vanish from Eq.~(\ref{eq:master_ge_complete}) for a
specific choice of the phases of the adiabatic basis states.

Let us explicitly reformulate and assess the range of validity of
the master equation [see Eqs.~(\ref{eq:master_gg_complete}) and
(\ref{eq:master_ge_complete})]. The approximation of adiabatic rates
requires that the external drive does not change the Hamiltonian
governing the system on the time scale of the memory time of the
bath, i.e., we assume that $\tau_{\text{corr}} \ll 1/||\hat{w}||$
where $\tau_{\text{corr}}$ is the environment correlation time.
Furthermore, we require that the dynamics of the density opertor
occur on time scales longer than the environment autocorrelation
time so that $\tau_{\text{corr}} \ll 1/\omega_0$, where $1/\omega_0$
is a typical system transition time relating to the off-diagonal
elements of the density matrix. As mentioned in Appendix, we apply
the Markov approximation by assuming that $\tau_{\text{corr}} \ll
1/\gamma$ where $1/\gamma$ is a typical relaxation time of the
system. Moreover, the assumptions we employ demand that $1/\omega_0
\ll 1/\gamma, 1/||\hat{w}||$. Combining the used approximations
leads to a requirement of time scale separation $\tau_{\text{corr}}
\ll 1/\omega_0 \ll 1/\gamma, 1/||\hat{w}||$.

Remarkably, our master equation is identical to that derived in Ref.~\cite{general_master}, however, the manner in which the master equation was derived is different. In Ref.~\cite{general_master}, one starts from the effective Hamiltonian for the adiabatic basis presented in Eq.~(\ref{eq:Heff}) and formulates a nested commutator expression for the derivative of the reduced system density operator in the adiabatic basis applying $\hbar \hat{w}(t) + \hat{\tilde{V}}(t)$ as the perturbation
\begin{widetext}
\begin{equation}
\begin{split}
\frac{d \hat{\tilde{\sigma}}_I(t)}{dt} &= i[\hat{\tilde{\sigma}}_I(t),\hat{w}_I(t)] -\frac{1}{\hbar^2} \textrm{Tr}_E \left\{  \int_0^t \, dt' \left[[\hat{\tilde{\rho}}_I(t),\hat{\tilde{V}}_I(t')],\hat{\tilde{V}}_I(t)\right] \right\} \\ &+ \frac{i}{\hbar^2} \textrm{Tr}_E \left\{ \int_0^t \, dt' \int_{0}^{t'} \, dt'' \left[ \left[ \hat{\tilde{\rho}}_I(t), [\hat{w}_I(t'),\hat{\tilde{V}}_I(t'')] \right], \hat{\tilde{V}}_I(t) \right] \right\},
\end{split}
\label{eq:dsigmadtnew1}
\end{equation}
\end{widetext}
in the interaction picture. Using this operator directly, results in the same master equation to the one we obtained. Thus, we find that with respect to adiabatic temporal evolution it makes no difference whether one uses the effective Hamiltonian for the adiabatic basis and takes $\hbar \hat{w}(t)+\hat{\tilde{V}}(t)$ as the perturbation as was done in Refs.~\cite{master, general_master} or whether one uses our approach to express the effective Hamiltonian for the super-adiabatic basis assuming that the super-adiabatic correction is small, thus taking $\hat{\tilde{V}}^{(2)}(t)$ as the perturbation and writing the super-adiabatic basis states up to the linear order in $\alpha(t)$. Our discovery reaffirms that the super-adiabatic basis approximates the exact evolving state in the next order in $\alpha$ so that using only the bath coupling as the perturbation will result in describing the dynamics in the same order as the effective Hamiltonian for the adiabatic basis does. This result is an important consistency check for the master equation derived in Refs.~\cite{master, general_master}, see Eqs.~(\ref{eq:master_gg_complete}) and (\ref{eq:master_ge_complete}).

The original way \cite{master, general_master} of deriving the full
master equation can be extended to obtain master equations in higher
orders in $\alpha$ by applying the nesting procedure iteratively
[see Eq.~(\ref{eq:dsigmadtnew1})]. Our method can be used as well to
obtain higher-order master equations by using higher-order
perturbation theory to write the matrix elements required to utilize
the master equation for non-steered systems. Since our technique is
based on applying algebraic operations, it is potentially simpler to
obtain higher order equations with it than with the nesting
procedure which results in complicated integral expressions.

As an example, let us illustrate how to obtain the master equation
up to the second order in $\alpha(t)$ using our method for a
two-level system. We begin by defining a unitary transformation
$\hat{D}_2(t)$ making $\hat{\tilde{H}}^{(2)}_S(t) + \hbar
\hat{w}_1(t)$ diagonal in the fixed basis $\{ \ket{0}, \ket{1}\}$.
Thus the relevant density matrix becomes $\hat{\tilde{\rho}}^{(3)} =
\hat{D}_2^{\dagger}\hat{\tilde{\rho}}^{(2)}\hat{D}_2$ and its
evolution is governed by the effective Hamiltonian for the second
super-adiabatic basis
\begin{equation}
\hat{\tilde{H}}^{(3)}(t) = \hat{\tilde{H}}^{(3)}_S(t) + \hbar \hat{w}_2(t) + \hat{\tilde{V}}^{(3)}(t) + \hat{H}_E,
\label{eq:Heff2}
\end{equation}
where $\hat{\tilde{H}}^{(3)}_S(t)=\hat{D}_2^{\dagger}(t)[\hat{\tilde{H}}_S^{(2)}(t)+\hbar \hat{w}_1(t)]\hat{D}_2(t)$, $\hat{\tilde{V}}^{(3)}(t)=\hat{D}_2^{\dagger}(t)\hat{\tilde{V}}^{(2)}(t)\hat{D}_2(t)$, and $\hat{w}_2=-i\hat{D}_2^{\dagger}(t)\dot{\hat{D}}_2(t)$. Assuming that $\hat{w}_2$ is negligible with respect to $\hat{w}_1$, we can again resort to the approach of Sec.~\ref{sec:non-steered} and write a master equation similar to Eqs.~(\ref{eq:rho_gg1}) and (\ref{eq:rho_ge}). Finding the relevant matrix elements is slightly more complicated than in the case of the master equation up to the linear order in $\alpha(t)$. We essentially wish to describe $\hat{D}(t)\hat{D}_1(t)\hat{D}_2(t)\ket{m}$, where $\ket{m}$ is a fixed state, up to the second order in $\alpha(t)$. We first write the first super-adiabatic eigenstates up to the second order in $\alpha(t)$ as
\begin{equation}
\ket{g^{(3)}} = \ket{g}-\ket{e}\frac{w_{ge}^*}{\omega_{01}} + \ket{e}\frac{w_{ee}w_{ge}^*}{\omega_{01}^2},
\label{eq:eig_corrected2_g}
\end{equation}
and
\begin{equation}
\ket{e^{(3)}} = \ket{e}+\ket{g}\frac{w_{ge}}{\omega_{01}} + \ket{g}\frac{w_{gg}w_{ge}}{\omega_{01}^2},
\label{eq:eig_corrected2_e}
\end{equation}
with the eigenenergies $E_g^{(3)}=E_g+\hbar w_{gg} - \hbar |w_{ge}|^2/\omega_{01}$ and $E_e^{(3)}=E_e+\hbar w_{ee} + \hbar |w_{ge}|^2/\omega_{01}$, respectively. Additionally, we must account for the lowest order in $||\hat{w}_1(t)|| \sim \alpha(t)^2$ to obtain the states of interest up to the second order in $\alpha(t)$ as
\begin{equation}
\hat{D}(t)\hat{D}_1(t)\hat{D}_2(t)\ket{0} \approx \ket{g^{(3)}} -
\ket{e}\frac{w_{1,ge}^*}{\omega_{01}}, \label{eq:eig_g_final}
\end{equation}
and
\begin{equation}
\hat{D}(t)\hat{D}_1(t)\hat{D}_2(t)\ket{1} \approx \ket{e^{(3)}} +
\ket{g}\frac{w_{1,ge}}{\omega_{01}}, \label{eq:eig_e_final}
\end{equation}
and the energy gap as $\omega_{01}^{(3)} = \omega_{01}^{(2)} + 2|w_{ge}|^2/\omega_{01} + (w_{1,ee}-w_{1,gg})$. Here we denote $w_{1,gg} = -i\braket{0|\hat{D}_1^{\dagger}(t)\dot{\hat{D}}_1(t)|0}$, $w_{1,ee} = -i\braket{1|\hat{D}_1^{\dagger}(t)\dot{\hat{D}}_1(t)|1}$ and $w_{1,ge} = -i\braket{0|\hat{D}_1^{\dagger}(t)\dot{\hat{D}}_1(t)|1}$. Using these definitions, one could write the matrix elements in the relevant master equation for a non-steered system to obtain the master equation for adiabatic steering up to the second order in $\alpha(t)$.

\section{\label{sec:opt_phase}Optimal phase selection}

Assume that we have nondegenerate adiabatic eigenstates $\ket{g}$
and $\ket{e}$ that are normalized and smooth during the temporal
evolution, and that they can be obtained from the fixed states with
a unitary transformation as $\hat{D}\ket{0}=\ket{g}$ and
$\hat{D}\ket{1}=\ket{e}$. Thus, the operator determining the
adiabatic evolution becomes
$\hat{w}=-i\hat{D}^{\dagger}\dot{\hat{D}}$ [see
Eq.~(\ref{eq:Heff})]. However, the choice for the complex phases of
the states is essentially arbitrary, and hence its effect on
$\hat{w}$ should be studied. This means that we could also work with
basis states that differ from $\ket{g}$ and $\ket{e}$ by phase
factors which depend on the point in the control cycle. In this
section, we show that this degree of freedom can be used to minimize
the local adiabatic parameter, while in the corresponding master
equation it leads only to renormalized matrix elements. Furthermore,
we discuss how to account for the noncyclic geometric phase leading
to a slightly different but generally more feasible phase selection.

Let us choose new phases for the states by multiplying them with phase factors $e^{i\lambda_g}$ and $e^{i\lambda_e}$, where $\lambda_g,\lambda_e \in \mathbb{R}$, so that a phase selection operator $\hat{\Omega}$ is defined as
\begin{equation}
\hat{\Omega} = e^{i\lambda_g}\ket{0}\bra{0} + e^{i\lambda_e}\ket{1}\bra{1},
\label{eq:omegaperm}
\end{equation}
yielding the new transformation as $\hat{\tilde{D}} = \hat{D}\hat{\Omega}$. Notice that the new states defined by the transformation are still eigenstates of the original Hamiltonian. With this transformation, the operator for the drive becomes
\begin{equation}
\begin{split}
\hat{\tilde{w}} &= -i \hat{\tilde{D}}^{\dagger}\dot{\hat{\tilde{D}}} \\ &= -i\hat{\Omega}^{\dagger}\hat{D}^{\dagger}(\dot{\hat{D}}\hat{\Omega}+\hat{D}\dot{\hat{\Omega}}) \\ &= \dot{\lambda}_g\ket{0}\bra{0} + \dot{\lambda}_e\ket{1}\bra{1} - i\hat{\Omega}^{\dagger}\hat{D}^{\dagger}\dot{\hat{D}}\hat{\Omega},
\end{split}
\label{eq:wnew}
\end{equation}
where we have used the unitarity of $\hat{D}$. The matrix elements in the phase shifted basis become
\begin{equation}
\begin{split}
\braket{0|\hat{\tilde{w}}|0} &= \dot{\lambda}_g-i\braket{0|\hat{D}^{\dagger}\dot{\hat{D}}|0} = \dot{\lambda}_g + w_{gg}, \\
\braket{1|\hat{\tilde{w}}|1} &= \dot{\lambda}_e-i\braket{1|\hat{D}^{\dagger}\dot{\hat{D}}|1} = \dot{\lambda}_e + w_{ee}, \\
\braket{0|\hat{\tilde{w}}|1} &= -i e^{i(\lambda_e-\lambda_g)}\braket{0|\hat{D}^{\dagger}\dot{\hat{D}}|1} = e^{i(\lambda_e-\lambda_g)}w_{ge}, \\
\braket{1|\hat{\tilde{w}}|0} &= -i e^{i(\lambda_g-\lambda_e)}\braket{1|\hat{D}^{\dagger}\dot{\hat{D}}|0} = e^{i(\lambda_g-\lambda_e)}w_{eg}.
\end{split}
\label{eq:wnewelements}
\end{equation}
Thus, the phase shift induces a shift in the diagonal elements and a phase shift in the off-diagonal elements.

Since the choice of the phases of the eigenstates changes the
$\hat{w}$ terms of the master equation and affects the quantum
evolution, it has to be fixed by physical reasoning. One way to
avoid artefact effects arising from the phase choice is to minimize
the Hilbert-Schmidt norm $||\hat{\tilde{w}}|| =
\sqrt{\mathrm{Tr}_S\{\hat{\tilde{w}}^{\dagger}\hat{\tilde{w}}\}}$ at
each time instant.
For this task,
it suffices to minimize
\begin{equation}
\begin{split}
\mathrm{Tr}_S \{ \hat{\tilde{w}}^{\dagger}\hat{\tilde{w}} \} &= |\braket{0|\hat{\tilde{w}}|0}|^2 + |\braket{1|\hat{\tilde{w}}|1}|^2 \\ &+ |\braket{1|\hat{\tilde{w}}|0}|^2 + |\braket{0|\hat{\tilde{w}}|1}|^2.
\end{split}
\label{eq:HSw}
\end{equation}
The last two terms consist of the off-diagonal terms and the phase selection has no effect on them. Thus, the minimum is found when we select the diagonal elements in Eq.~(\ref{eq:HSw}) to vanish, yielding
\begin{equation}
\begin{split}
\lambda_g(t) &= - \int_0^t dt' w_{gg}(t') + \lambda_g^0 = i \int_0^t dt' \braket{g|\dot{g}} + \lambda_g^0, \\
\lambda_e(t) &= - \int_0^t dt' w_{ee}(t') + \lambda_e^0 = i \int_0^t dt' \braket{e|\dot{e}} + \lambda_e^0,
\end{split}
\label{eq:lambdas}
\end{equation}
and
\begin{equation}
\begin{split}
\braket{0|\hat{\tilde{w}}|1} &= e^{i(\lambda_e^0-\lambda_g^0)}e^{i\int_0^t dt' [w_{gg}(t')-w_{ee}(t')]}w_{ge} \\ &= e^{i(\lambda_e^0-\lambda_g^0)}e^{\int_0^t dt' [\braket{g|\dot{g}}-\braket{e|\dot{e}}]}w_{ge}.
\end{split}
\label{eq:newoff}
\end{equation}
The absolute phases are not fixed since we have a degree of freedom
in the selection of the constant parts $\lambda_e^0$ and
$\lambda_g^0$. Notice that the primary selection of the smooth
eigenstates $\ket{g}$ and $\ket{e}$ determines the accumulating
phase. We denote the integrals as simply over time, but one should
bear in mind that they contribute a path in the parameter space.
Used in conjunction with our master equation, the above phase
selection results in $w_{gg}$ and $w_{ee}$ vanishing in
Eq.~(\ref{eq:master_ge_complete}). Furthermore, it minimizes the
local adiabatic parameter
$\tilde{\alpha}(t)=||\hat{\tilde{w}}(t)||/\omega_{01}(t)$. Thus we
refer to it as \emph{optimal phase selection} although it may not
yield the most accurate evolution as discussed below.

Utilizing the optimal phase selection scheme with our master
equation requires a careful consideration of the used
approximations, in particular, the approximation of adiabatic rates.
We used this approximation in the derivation of the master equation
[see Eqs.~(\ref{eq:master_gg_complete}) and
(\ref{eq:master_ge_complete})] to state that $\omega_{01}$, $m_1$,
$m_2$, and the matrix elements of $\hat{w}$ vary slowly in time.
With the optimal phase selection, the corresponding parameters are
$\tilde{\omega}_{01} = \omega_{01}$, $\tilde{m}_1 = m_1$,
$\tilde{m}_2 = e^{i(\lambda_e-\lambda_g)}m_2$,
$\tilde{w}_{gg}=\tilde{w}_{ee}=0$, and $\tilde{w}_{ge} =
e^{i(\lambda_e-\lambda_g)}w_{ge}$. Since the approximation of
adiabatic rates applies for the derivatives of the accumulating
phases defined in Eq.~(\ref{eq:lambdas}), any possible shift in the
power spectra induced by the optimal phase selection can be
neglected since it would only lead to higher order terms in the
master equation. Thus, the master equation can be directly used by
replacing the original variables with the phase shifted ones.

With the optimal selection, the phase shifted basis states become $\ket{\tilde{g}} = e^{i\lambda_g^0}e^{- \int_0^t dt' \braket{g|\dot{g}}}\ket{g}$ and $\ket{\tilde{e}} = e^{i\lambda_e^0}e^{- \int_0^t dt' \braket{e|\dot{e}}}\ket{e}$. Thus we have $\braket{\tilde{g}|\dot{\tilde{g}}} = \braket{\tilde{e}|\dot{\tilde{e}}} = 0$ independent of $\ket{g}$ and $\ket{e}$ and the selection renders the phase shifted states to be also invariant under a local gauge change, i.e., $\ket{g} \rightarrow e^{i\beta(t)}\ket{g}$ has no effect on $\ket{\tilde{g}}$ and $\ket{e} \rightarrow e^{i\eta(t)}\ket{e}$ has no effect on $\ket{\tilde{e}}$ where $\beta(t)$ and $\eta(t)$ are any smooth functions. For a closed path in the parameter space $\gamma$, we have
\begin{equation}
\begin{split}
\lambda_g(t_b) - \lambda_g(t_a) &= i \oint_{\gamma} \braket{g|\dot{g}},  \\
\lambda_e(t_b) - \lambda_e(t_a) &= i \oint_{\gamma} \braket{e|\dot{e}},
\end{split}
\label{eq:phaseberry}
\end{equation}
where we have denoted $t_a$ and $t_b$ as the virtual starting and
ending time instants for the path, respectively. These are the Berry
phases accumulated over the path for the phase shifted basis states
and, as such, cannot be removed by any continuous local gauge change
\cite{prsla392/45}. Thus, selecting the optimal local phase for a
closed loop in the parameter space implies a gauge-invariant
accumulated phase at the end of the loop. However, the optimal phase selection
neglects the effects of the accumulation speed of the geometric
phase on the environment-induced transitions in the master equation.
Although this effect is negligible in the adiabatic limit, it is
useful to provide means to take it into account consistently.

To go beyond the optimal phase selection, we need to study the
adiabatic noncyclic geometric phase acquired during the
evolution~\cite{prl60/2339, ajp66/431}. The adiabatic noncyclic
geometric phase acquired by the $n$th eigenstate~\cite{ajp66/431}
can be defined as
\begin{equation}
\begin{split}
\gamma_n(t) &= \arg \{\braket{n;\vec{q}(t_0)|n;\vec{q}(t)}\} \\ &+
i\int_0^t d\tau\braket{n;\vec{q}(\tau)|\partial_{\tau}|
n;\vec{q}(\tau)},
\end{split}
\label{eq:non-cyclic_phase}
\end{equation}
where $\ket{n;\vec{q}(t_0)}$ is an essentially arbitrary reference
eigenvector not orthogonal to $\ket{n;\vec{q}(t)}$. The phase
$\gamma_n(t)$ is invariant under any local phase transformation of
the basis states and independent of the speed, at which we traverse
the open path in the control parameter space. To allow for the
accumulating phase of the basis vectors to describe the noncyclic
geometric phase, one can define a phase transformation of the
eigenstates $\ket{n} = e^{-i\arg
\{\braket{n';\vec{q}(t_0)|n';\vec{q}(t)}\}}\ket{n'}$ where $n\in
\{g,e\}$ and $\ket{n'}$ describes an eigenstate with an arbitrary
continuous local phase. This transformation produces
\begin{equation}
\begin{split}
w_{nn} &= -\partial_t \arg
\{\braket{n';\vec{q}(t_0)|n';\vec{q}(t)}\} - i\braket{n'|\dot{n'}}
\\ &= -\dot{\gamma}_n(t).
\end{split}
\label{eq:non-cyclic_w}
\end{equation}
Thus the time-local accumulation speed of the geometric phase can be
made to appear in the master equation
[Eqs.~(\ref{eq:master_gg_complete}) and
(\ref{eq:master_ge_complete})]. However, the accumulation speed in
Eq.~(\ref{eq:non-cyclic_w}) is dependent on the reference point
$\ket{n';\vec{q}(t_0)}$ and, thus, redefining the reference changes
typically the accumulation speed at all times. This can become a
problem since the accumulation speed affects in general the
resulting physical quantities which should not depend on the choice
of the reference point. A possible way to correct for this
inconsistency is to use the geodesic approach \cite{prl60/2339}.

\section{\label{sec:conclusions}Conclusions}

We deviced a way to derive the full master equation for adiabatically steered quantum systems in the two-state approximation under the influence of decoherence starting from an interaction-picture-based derivation, in which the external drive was first omitted. The full master equation was obtained by approximating the transformation to the superadiabatic basis using the perturbation theory and exploiting the master equation for the non-steered system. We showed that the master equation we obtain this way is the same as the one obtained in Ref.~\cite{general_master} by a longer calculation. We concluded that our manner of obtaining the master equation is a consequence of the superadiabatic basis approximating the exact evolving state in the linear order in the adiabatic parameter $\alpha(t)$. Furthermore, there is no need to evaluate high-order nested commutators of integrals in our method if it is extended beyond the linear order in $\alpha(t)$ as opposed to the method in Refs.~\cite{master, general_master}. A detailed study of the efficiency of these two approaches is left for future research.

There exists a gauge degree of freedom in the choice of the phases
of the basis states during the evolution. We demonstrated a way to
choose the phases in a way which minimizes the local adiabatic
parameter and simplifies the derived master equation. We showed that
this choice produces basis states which are invariant under a local
gauge change.
Furthermore, we discussed how to account in a gauge invariant manner
for the effects of the accumulation speed of the adiabatic noncyclic
geometric phase on the environment-induced transitions.

\begin{acknowledgments}
The authors thank J.~P.~Pekola and E.~Sjöqvist for stimulating
discussions. We acknowledge Academy of Finland and Emil Aaltonen
Foundation for financial support. We have received funding from the
European Community's Seventh Framework Programme under Grant
Agreement No. 238345 (GEOMDISS).
\end{acknowledgments}

\appendix*

\section{\label{sec:appendix}Non-steered master equation in the two-state basis}

The reduced system density matrix in the interaction picture
$\hat{\sigma}_I(t) = \textrm{Tr}_E\{\hat{\rho}_I(t)\}$ can be used
to derive the relevant master equation assuming a stationary
environment, i.e., $\frac{d\hat{\rho}_E}{dt} =
\frac{i}{\hbar}[\hat{\rho}_E,\hat{H}_E] = 0$. We define the
operators in the interaction picture as $\hat{Z}_I(t) =
e^{i\hat{H}_Et/\hbar} \hat{U}_S^{\dagger}(t,0)\hat{Z}(t)
\hat{U}_S(t,0) e^{-i\hat{H}_Et/\hbar}$, where $\hat{Z}(t)$ is the
operator in the Schrödinger picture and $\hat{U}_S(t,0)$ is the
time-evolution operator. For a time dependent system Hamiltonian,
the time-evolution operator is $\hat{U}_S(t,0) =
\mathcal{T}e^{-i\int_0^t \hat{H}_S(\tau)d\tau/\hbar}$ but simplifies
to $\hat{U}_S(t,0) = e^{-i\hat{H}_St/\hbar}$ for non-steered systems
studied in this Appendix.

If we assume that the system interacts weakly with the environment, the master equation acquires the standard form (Redfield equation \cite{tToOQS})
\begin{equation}
\frac{d \hat{\sigma}_I(t)}{dt} = - \frac{1}{\hbar^2} \int_0^t \, dt' \mathrm{Tr}_E ([[\hat{\sigma}_I(t) \otimes \hat{\rho}_E,\hat{V}_I(t')],\hat{V}_I(t)]),
\label{eq:APPdsigmadt_final}
\end{equation}
in the interaction picture, where we have utilized the Born-Markov approximation \cite{tToOQS}. The transformation from the interaction picture to the Schrödinger picture unfolds when we employ
\begin{equation}
\hat{\rho}_S(t) = \hat{U}_S(t,0) \hat{\sigma}_I(t) \hat{U}_S^{\dagger}(t,0),
\label{eq:APPsigma_trans}
\end{equation}
which can be used to obtain the density matrix transformation componentwise as
\begin{equation}
\begin{split}
\rho_{gg}(t) &= \sigma_{I,gg}(t), \\
\rho_{ee}(t) &= \sigma_{I,ee}(t), \\
\rho_{ge}(t) &= e^{i\omega_{01}t}\sigma_{I,ge}(t), \\
\rho_{eg}(t) &= e^{-i\omega_{01}t}\sigma_{I,eg}(t).
\end{split}
\label{eq:APPdensity_matrix_trans}
\end{equation}
Derivating Eq.~(\ref{eq:APPsigma_trans}) yields the transformation of the derivative as
\begin{equation}
\frac{d \hat{\rho}_S(t)}{dt} = \frac{i}{\hbar} [\hat{\rho}_S(t),\hat{H}_S(t)] + \hat{U}_S(t,0) \frac{d \hat{\sigma}_I(t)}{dt} \hat{U}_S^{\dagger}(t,0).
\label{eq:APPderivative_trans}
\end{equation}
Using Eqs.~(\ref{eq:APPdsigmadt_final}) and (\ref{eq:APPderivative_trans}), we define the diagonal matrix element
\begin{widetext}
\begin{equation}
\begin{split}
\braket{g|\frac{d \hat{\rho}_S(t)}{dt}|g} = -\frac{1}{\hbar^2} \int_0^t \, dt' \textrm{Tr}_E\{ \braket{g|[[\hat{\sigma}_I(t)\otimes \hat{\rho}_E,\hat{V}_I(t')],\hat{V}_I(t)]|g}\},
\end{split}
\label{eq:APPbraket_der_diag}
\end{equation}
\end{widetext}
and the off-diagonal matrix element
\begin{widetext}
\begin{equation}
\begin{split}
\braket{g|\frac{d \hat{\rho}_S(t)}{dt}|e} = i\omega_{01}\rho_{ge}(t) - e^{i\omega_{01}t} \frac{1}{\hbar^2} \int_0^t \, dt' \textrm{Tr}_E\{ \braket{g|[[\hat{\sigma}_I(t)\otimes \hat{\rho}_E,\hat{V}_I(t')],\hat{V}_I(t)]|e}\},
\end{split}
\label{eq:APPbraket_der_off-diag}
\end{equation}
\end{widetext}
for the non-steered master equation. Notice that our derivation is
based on assuming that the system relaxation time is long compared
to the environment correlation time $\tau_{\mathrm{corr}}$ so that
the environment has no memory, i.e., we are in the Markov regime
\cite{API}. This allows us to neglect any variation of
$\hat{\sigma}_I(t)$ between times $t$ and $t+\tau_{\mathrm{corr}}$.
The integral expressions in Eqs.~(\ref{eq:APPbraket_der_diag}) and
(\ref{eq:APPbraket_der_off-diag}) simplify to give
Eqs.~(\ref{eq:rho_gg1}) and (\ref{eq:rho_ge}) when we expand the
commutators, use the closure relation for the adiabatic basis, and
utilize $\textrm{Tr}_E \{\hat{\rho}_E \hat{X}(t') \hat{X}(t)\} =
\int_{-\infty}^{\infty} \frac{d\omega}{2\pi} S_X(\omega)e^{-i\omega
(t'-t)}$ and Eq.~(\ref{eq:APPdensity_matrix_trans}). Furthermore, we
assume that the system time scales are longer than the system
autocorrelation time to approximate the spectral densities in the
remaining integral expressions. This treatment leads to neglecting
the Lamb shift.

\bibliography{localbib}

\end{document}